\documentclass[twoside]{article}

\usepackage{graphicx,rotating,subfigure,amsmath,amsfonts,amssymb,delarray}

\usepackage[T2A]{fontenc}
\usepackage[cp1251]{inputenc}
\usepackage{cmpj2e}

\usepackage{citesort}

\newcommand{\e}{\text{e}}
\newcommand{\im}{\text{i}}
\newcommand{\bulk}{{\mbox{\tiny bulk}}}
\def\l{\left}
\def\r{\right}
\def\12{\frac{1}{2}}
\def\wt{\widetilde}
\def\nn{\nonumber}

\title[Spin-$1/2$ chain compounds]{Consequences of lattice imperfections and interchain couplings for the
  critical properties of spin-$1/2$ chain compounds}
\author{J. Sirker}
\address{Max-Planck-Institute for Solid State Research, D-70569 Stuttgart,
  Germany }
\makeatletter
\newif\if@eqnobysec
\def\theequation{\if@eqnobysec
      \arabic{section}.\arabic{equation}\else
      \arabic{equation}\fi}
\def\eqnobysec{\@eqnobysectrue\@addtoreset{equation}{section}}
\def\eqnotbysec{\@eqnobysecfalse}
\makeatother

\eqnobysec
\begin{document}

\maketitle

\begin{abstract}
  To allow for a comparison of theoretical predictions for spin chains with
  experimental data, it is often important to take impurity effects as well as
  interchain couplings into account. Here we present the field theory for
  finite spin chains at finite temperature and calculate experimentally
  measurable quantities like susceptibilities and nuclear magnetic resonance
  spectra. For the interchain couplings we concentrate on geometries relevant
  for cuprate spin chains like Sr$_2$CuO$_3$ and SrCuO$_2$. The field
  theoretical results are compared to experimental as well as numerical data
  obtained by the density matrix renormalization group.
\pacs 75.10.Pq, 75.10.Jm, 11.10.Wx, 02.30.Ik
\end{abstract}

\section{Introduction}
\label{Intro}
A large number of materials are known which, over a certain temperature range,
are well described by simple spin chain or ladder models
\cite{MotoyamaEisaki,PrattBlundell,ThurberHunt,SologubenkoGianno,Kojima,TakigawaMotoyama,BoucherTakigawa,RosnerEschrig,Hase,MaBroholm,RueggKiefer}.
In all these Mott insulators the superexchange constants are spatially very
anisotropic so that a three-dimensional crystal effectively shows
one-dimensional magnetic properties. In one of the best known spin-$1/2$ chain
compounds Sr$_2$CuO$_3$, for example, the superexchange constant along the
chain direction $J\sim 2200$ K whereas the magnetic couplings $J_\perp$ along
the other directions are at least three orders of magnitude smaller
\cite{MotoyamaEisaki,RosnerEschrig}. In such a system one can therefore
experimentally explore the physical properties of a spin chain over a very
wide temperature range $J_\perp \ll T \lesssim J$. From a theoretical
perspective this is very exciting because it allows to study experimentally
many aspects of one-dimensional field theories
\cite{egg94,EggertAffleck92,GiamarchiBook}. In addition, the ideal Heisenberg
spin-$1/2$ chain is integrable so that compounds well described by this model
make it possible to experimentally address the question how physical
properties, in particular, transport, are affected by a nearby integrable
point \cite{ThurberHunt,PrattBlundell,SologubenkoGianno}. Here the infinite
set of constants of motion making the model integrable is expected to slow
down the decay of current correlations - or even prevent them from decaying
completely - leading to anomalous transport properties \cite{ZotosPrelovsek}.

In real materials, however, we are always confronted with impurities and
lattice imperfections which weaken or even completely destroy a superexchange
bond between two spins. Because a weakening of a bond is a relevant
perturbation in the renormalization group (RG) sense, we have to deal - at
least at low temperatures - with finite chains with open boundary conditions
(OBCs). Measurements on such systems then correspond to taking averages over
ensembles of finite chain segments with lengths determined by a distribution
function \cite{SirkerLaflorencie,SirkerLaflorencie2}. Furthermore, the
description by a one dimensional model will break down for temperatures $T\sim
J_\perp$ where usually a three-dimensional magnetic order sets in. However,
even for temperatures above this ordering temperature interchain couplings can
have a significant effect which has to be taken into account.

In the following, we will study the anisotropic spin-$1/2$ Heisenberg chain
(XXZ model) with $N$ sites and OBCs
\begin{equation}
\label{eq1}
H = J\sum_{j=1}^{N-1} \l[ S^x_j S^x_{j+1} + S^y_j S^y_{j+1} + \Delta S^z_j
S^z_{j+1} \r] - h\sum_{j=1}^N S^z_j \; .
\end{equation}
Here $J$ is the exchange constant and $h$ the applied magnetic field.
Although exchange anisotropies due to spin-orbit coupling are usually rather
small so that experimentally only the isotropic case $\Delta=1$ is relevant,
the additional parameter $\Delta$ is useful for the field theoretical
calculations in section \ref{FT}. There are two important consequences of the
OBCs.  First, expectation values of local operators become position dependent
because translational invariance is broken. In the following we will in
particular study the {\it local susceptibility} defined as
 \begin{equation}
\label{eq2} 
\chi_j=\frac{\partial}{\partial h} \langle S^z_j\rangle_{h=0} = \frac{1}{T} \langle
S^z_j S^z_{\rm tot}\rangle_{h=0}
\end{equation}
where $S^z_{\rm tot} = \sum_j S^z_j$. Second, there are well defined boundary
contributions to all thermodynamic quantities. For $N\to \infty$ the total
free energy $F$ is, for example, given by
\begin{equation}
\label{eq3} 
F = Nf_{\rm bulk} + F_B +\mathcal{O}(1/N) \; , 
\end{equation}
where $F_B$ is the boundary free energy. Similarly, one can define a boundary
susceptibility $\chi_B$
\cite{FujimotoEggert,FurusakiHikihara,BortzSirker,SirkerBortzJSTAT,SirkerBortz}. These
boundary or surface terms will be studied here as well. The local
susceptibility defined in Eq.~(\ref{eq2}) can be related to the boundary
susceptibility by
 \begin{equation}
\label{eq4} 
\chi_B = \lim_{N\to\infty}\l(\sum_{j=1}^N\chi_j - N\chi_\bulk\r)
\end{equation}
where $\chi_{\bulk}$ is the bulk susceptibility defined analogously to the
bulk free energy in Eq.~(\ref{eq3}).

\begin{figure}[t!]
\begin{center}
\includegraphics*[width=0.7\columnwidth]{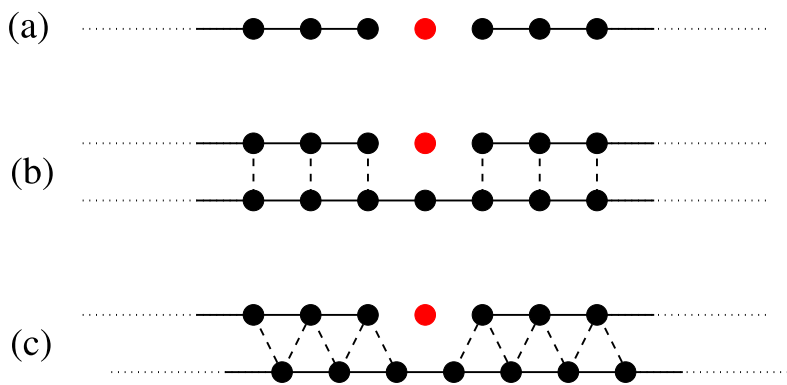}
\end{center}
\caption{A non-magnetic impurity in (a) a single spin chain, (b) in a spin
  chain coupled to a neighboring chain by a ladder-like interchain coupling,
  and (c) in a spin chain coupled in a zigzag fashion to a neighboring chain.}
\label{fig_geom}
\end{figure}

For the interchain couplings we will consider two different cases relevant for
many materials. One is a simple ladder-like antiferromagnetic coupling between
neighboring chains as, for example, in Sr$_2$CuO$_3$ but also many other spin
chain compounds. This case is shown in Fig.~\ref{fig_geom}(b). The other is a
zigzag {\it ferromagnetic} coupling between neighboring chains. This kind of
interchain coupling is sketched in Fig.~\ref{fig_geom}(c) and is relevant, for
example, for SrCuO$_2$ \cite{MotoyamaEisaki}.

To provide an intuitive picture we start with some numerical results for
semi-infinite spin chains with geometries as shown in Fig.~\ref{fig_geom}
obtained by the density matrix renormalization group applied to transfer
matrices (TMRG) in section \ref{TMRG}. In section \ref{FT} we will then
present the field theory for finite Heisenberg chains with OBCs at finite
temperatures. In section \ref{Chi} we use the results obtained in the previous
section to calculate the susceptibility as an ensemble average over finite
spin chain segments and show that the obtained results are in good agreement
with experimental measurements. In section \ref{KS} we derive in a similar
fashion the nuclear magnetic resonance (NMR) spectrum and show that it
provides information about the interchain couplings. Finally, we give a brief
summary and present some conclusions.

\section{Numerical results}
\label{TMRG}
A method particularly suited to calculate the thermodynamic properties of
one-dimensional systems is the density matrix renormalization group applied to
transfer matrices (TMRG)
\cite{BursillGehring,WangXiang,Shibata,SirkerKLuemperPRB,SirkerKluemperEPL,SirkerKluemperDTMRG,BortzSirker,SirkerBortz,SirkerBortzJSTAT}.
To this end, the one-dimensional quantum system is mapped onto a
two-dimensional classical system by a Trotter-Suzuki decomposition
\cite{Trotter,Suzuki1,Suzuki2}. The additional dimension then corresponds to
the inverse temperature $\beta$. For the classical model a transfer matrix is
defined which evolves along the spatial direction. Importantly, one can show
that even for a critical system there is always a gap between the leading
eigenvalue $\Lambda_0$ and next-leading eigenvalues $\Lambda_\alpha$ of the
transfer matrix $T$ with $\xi_\alpha^{-1}=\ln|\Lambda_0/\Lambda_\alpha|$
defining a correlation length. This makes it possible to perform the
thermodynamic limit, i.e., system size $N\to\infty$, exactly. With the TMRG
one can treat impurity problems
\cite{EggertRommer,SirkerBortz,SirkerBortzJSTAT} as well as frustrated systems
\cite{Raupach} making it the ideal numerical tool to study the problem
considered here.

We will start by investigating the local susceptibility as defined in
Eq.~(\ref{eq2}) for a semi-infinite chain. By this we mean a chain which is
infinitely long but has one end with OBCs. To obtain $\chi_j$ we calculate the
local magnetization $\langle S^z_j\rangle$ for small magnetic fields $h/J\sim
10^{-2}$ and take a numerical derivative. Within the transfer matrix
formalism the local magnetization is given by
\begin{equation}
\label{TMRG.1} 
\lim_{N\rightarrow\infty}\langle S^z_j\rangle = \frac{\langle\Psi_L^0|T(S^z)T^{j-1}\wt{T}|\Psi_R^0\rangle}
{\Lambda_0^j \langle\Psi_L^0|\wt{T}|\Psi_R^0\rangle} \; .
\end{equation}
Here $\wt{T}$ is a modified transfer matrix containing the broken bond,
$T(S^z)$ is the transfer matrix with the operator $S^z$ included, and
$|\Psi_R^0\rangle$ ($\langle\Psi_L^0|$) the right (left) eigenstates belonging
to the largest eigenvalue $\Lambda_0$, respectively. Far away from the
boundary $\langle S^z_j\rangle$ becomes a constant, the bulk magnetization
\begin{eqnarray}
\label{TMRG.2}
m= \lim_{j\rightarrow\infty}\lim_{N\rightarrow\infty}\langle S^z_j\rangle 
&=& \lim_{j\rightarrow\infty}\frac{\sum_n\langle\Psi_L^0|T(S^z)T^{j-1}|\Psi_R^n\rangle\langle\Psi_L^n|\wt{T}|\Psi_R^0\rangle}
{\Lambda_0^j \langle\Psi_L^0|\wt{T}|\Psi_R^0\rangle} \nonumber \\
&=& \frac{\langle\Psi_L^0|T(S^z)|\Psi_R^0\rangle}{\Lambda_0}   \; .
\end{eqnarray}
By taking again a numerical derivative with respect to a small magnetic field,
we can obtain the bulk susceptibility $\chi_{\bulk}$ from (\ref{TMRG.2}).

\begin{figure}[t!]
  \begin{center}
\includegraphics*[width=0.7\columnwidth]{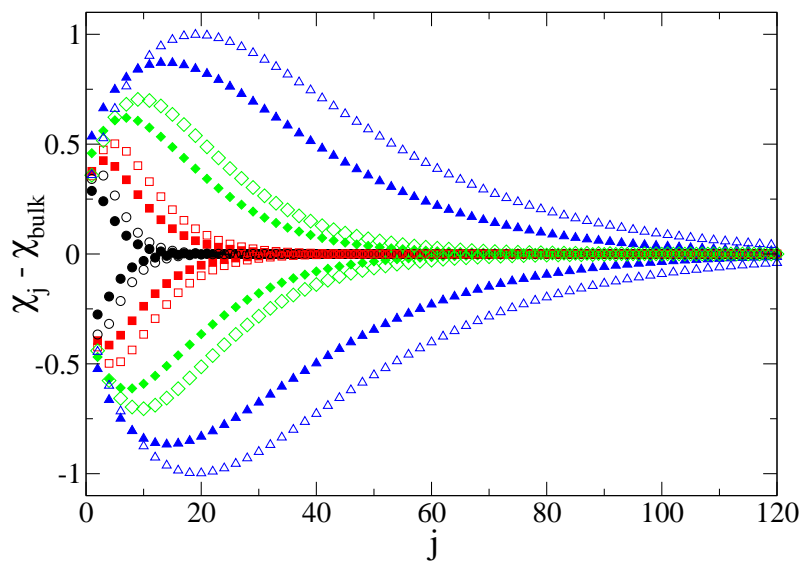}
\end{center}
\caption{$\chi_j-\chi_\bulk$ for a semi-infinite chain with $\Delta=1$ at
  temperatures $T/J=0.2,\, 0.1,\, 0.05,\, 0.025$. The numerical TMRG data
  (closed symbols) are compared to the field theory formula
  (\ref{StaggPart4_iso}) from section \ref{KS} (open symbols).}
\label{fig_chain}
\end{figure}

In Fig.~\ref{fig_chain} the susceptibility profile $\chi_j - \chi_{\rm bulk}$
for a single semi-infinite chain with $\Delta=1$ is shown.
As might be expected, the boundary induces Friedel-like oscillations which
become larger with decreasing temperature. Interestingly, at low temperatures
the oscillations first {\it increase} and reach a maximum, before decaying at
large distances. This phenomenon has been first studied by Eggert and Affleck
in \cite{EggertAffleck95} and we will rederive their field theory result as a
special case of our more general considerations in section \ref{FT}.

\begin{figure}[t!]
\begin{center}
  \includegraphics*[width=0.7\columnwidth]{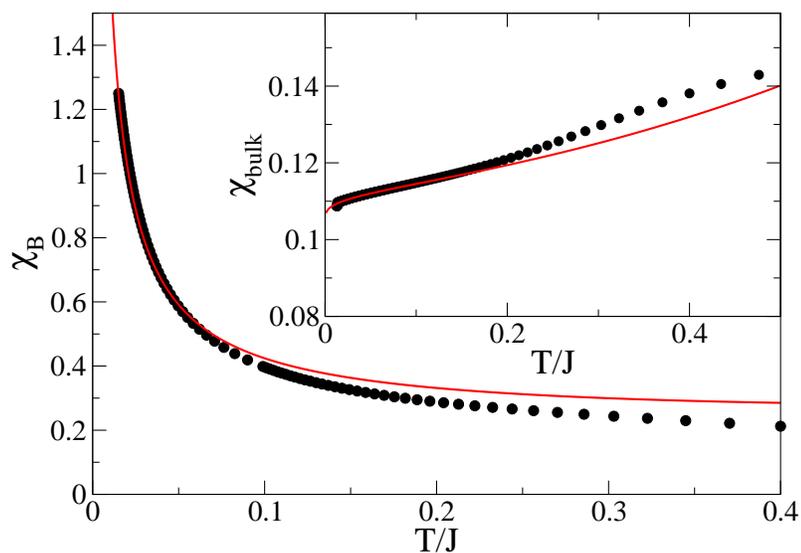}
\end{center}
\caption{The boundary susceptibility $\chi_B$ for $\Delta=1$ as obtained by
  TMRG (symbols) compared to the field theory result (\ref{chiB_iso3}) (line)
  derived in section \ref{FT} which is valid at low temperatures. Inset:
  Numerical results for the bulk susceptibility $\chi_{\bulk}$ for
  $\Delta=1$ (symbols) compared to the field theory formula
  (\ref{bulk_thermo}) (line).}
\label{fig_ChiB}
\end{figure}

Next, we consider the boundary susceptibility which we can easily obtain from
the susceptibility profile using Eq.~(\ref{eq4}). The result is shown in
Fig.~\ref{fig_ChiB}.
At low temperatures it was shown analytically that $\chi_B\sim
[T\ln(T_0/T)]^{-1}$ with a known constant $T_0$
\cite{FujimotoEggert,FurusakiHikihara,SirkerLaflorencie2}. The bulk
susceptibility, on the other hand, behaves as $\chi_{\rm bulk}\sim \mbox{const} +
\ln^{-1}(T_0/T)$ \cite{egg94}, i.e., it goes to finite value at $T=0$ with
infinite slope. We will come back to this in section \ref{FT} but already
notice that the numerical data are well described by the field theory.

\begin{figure}[t!]
  \begin{center}
\includegraphics*[width=0.7\columnwidth]{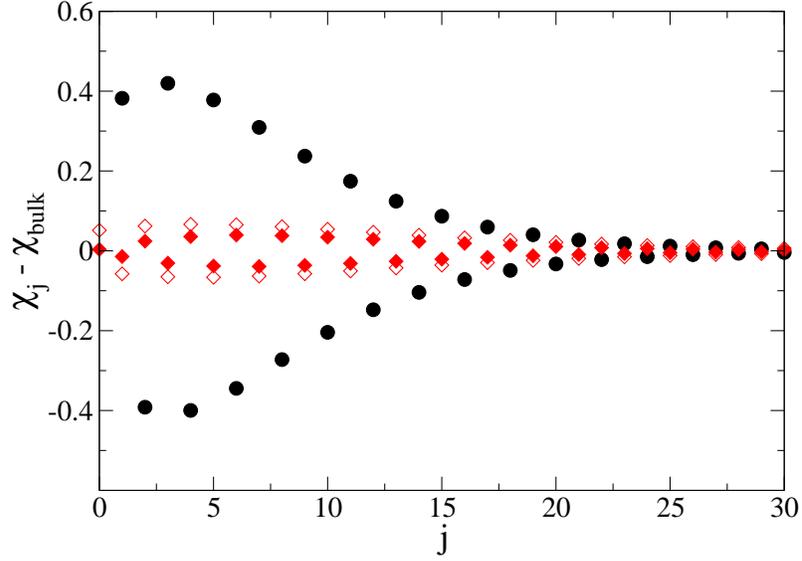}
\end{center}
\caption{$\chi_j-\chi_\bulk$ for $\Delta=1$, $J_\perp=0.03\, J$ and $T=0.09\,
  J$ and a ladder-like interchain coupling as shown in Fig.~\ref{fig_geom}(b).
  One of the chains has a non-magnetic impurity at site $j=0$ (circles) whereas
  the other one (diamonds) is infinitely long and does not have any
  impurities. The numerical TMRG data (closed symbols) are compared to the
  field theory formula (\ref{reflection}) from section \ref{KS} (open
  symbols).}
\label{fig_ladder}
\end{figure}

\begin{figure}[t!]
\begin{center}
  \includegraphics*[width=0.7\columnwidth]{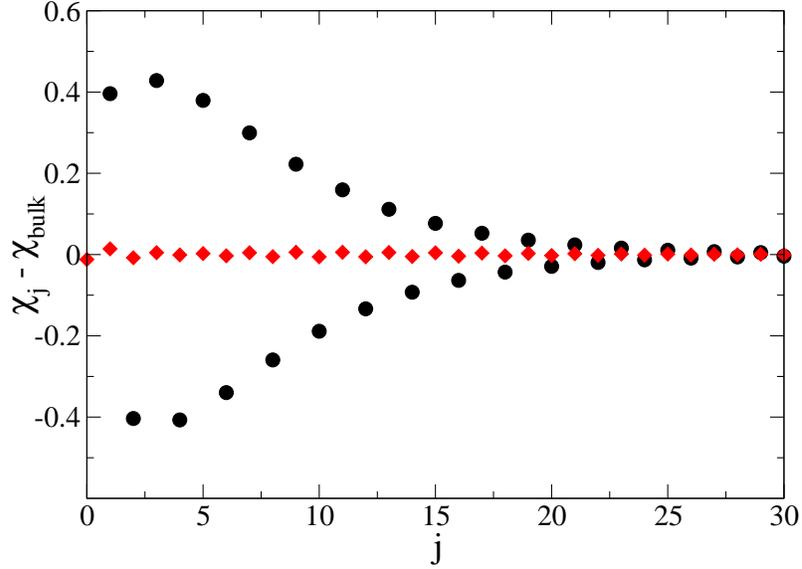}
\end{center}
\caption{$\chi_j-\chi_\bulk$ for $\Delta=1$, $J_\perp=-0.1\, J$ and $T=0.09\,
  J$ with a zigzag interchain coupling as shown in Fig.~\ref{fig_geom}(c). One
  of the chains has a non-magnetic impurity at site $j=0$ (circles) whereas the
  other one (diamonds) is infinitely long does not have any impurities. For
  this case field theory predicts that there are no
  oscillations, $\chi_j-\chi_\bulk\approx 0$, in the infinitely long chain.}
\label{fig_zigzag}
\end{figure}

Finally, we also have calculated susceptibility profiles for the cases of
coupled chains shown in Fig.~\ref{fig_geom}(b) and (c). A ladder-like coupling
of neighboring chains is, for example, realized in Sr$_2$CuO$_3$
\cite{MotoyamaEisaki}. In Fig.~\ref{fig_ladder} it is shown that in this case
an impurity in one chain also has a significant effect on a neighboring chain
without impurities. The Friedel-like oscillations are reflected in the
impurity-free chain due to the interchain couplings.
Clearly, the size of the reflected Friedel-like oscillations will depend on
the ratio $J_\perp/T$. However, this ratio is not the only relevant factor.
The geometry of the interchain couplings plays an important role as well. In
SrCuO$_2$ neighboring chains are coupled by a ferromagnetic zigzag-type
coupling. In this case an impurity in one chain leaves a neighboring chain
almost unaffected even if the temperature $T\sim J_\perp$ as is shown in
Fig.~\ref{fig_zigzag}.
In section \ref{KS} we will show that these differences can be easily
understood if one starts from the field theory results for a single chain
derived in the next section and takes the interchain couplings into account
perturbatively.

\section{Field theory for finite spin chains}
\label{FT}
In the limit of low temperatures and large chain length, the XXZ model
(\ref{eq1}) can be represented by a field theory. The main step is a
linearization of the dispersion around the two Fermi points. A certain linear
combination of particle-hole excitations around a Fermi point then constitutes
a collective bosonic mode described by the Luttinger liquid Hamiltonian
\begin{eqnarray}
  \label{FT.1}
  H \!&=& \!\frac{v}{2} \int_0^{L+a} \!\!\!\!\!\!\!\! dx \l[\Pi^2+(\partial_x\Phi)^2\r] -
  h\sqrt{\frac{K}{2\pi}}\!\int_0^{L+a} \!\!\!\!\!\!\!\! dx \,\partial_x\Phi 
\end{eqnarray}
where $v$ is the spin velocity, $L=Na$, and $a$ the lattice constant.  The
bosonic field $\Phi$ obeys the standard commutation rule
$[\Phi(x),\Pi(x')]=\im\delta(x-x')$ with $\Pi=v^{-1}\partial_t\Phi$. The
Luttinger parameter $K$ is a function of the anisotropy $\Delta$ and can be
determined exactly by Bethe ansatz with $K=1$ at the isotropic point. The spin
operators can be directly expressed in terms of the boson $\Phi$, in
particular, we have
 \begin{equation}
   \label{FT.2}
S^z_j \approx \sqrt{\frac{K}{2\pi}}\partial_x\Phi + c (-1)^j \cos\sqrt{2\pi
    K}\Phi 
\end{equation}
at zero magnetic field. Here $c$ is an amplitude which can also be obtained
exactly \cite{aff_corr,LukyanovTerras,SirkerLaflorencieNMR}. The separation of
$S^z_j$ into a uniform and a staggered part at low energies can be understood
as follows. In the equivalent spinless fermion representation of the XXZ
model, obtained by a Jordan-Wigner transformation, the $S^z$ operator becomes
the density operator. Due to the linearization of the dispersion the electrons
can live either at the same Fermi point and have therefore small momentum or
they can be situated at different Fermi points in which case the associated
momentum is $2k_F$. Zero magnetic field corresponds to half-filling for the
spinless fermions thus $2k_F=\pi$ and we obtain the staggered contribution in
(\ref{FT.2}). This also means that the local susceptibility defined in
Eq.~(\ref{eq2}) can be separated at low temperatures into a uniform and a
staggered part
 \begin{equation}
\label{FT.3}
\chi_j = \chi^{\rm uni} +  (-1)^j\chi_j^{\rm st} \; .
\end{equation}
In a bulk susceptibility measurement the staggered part does not contribute.
This part is, however, measurable in probes of the local magnetism as for
example in NMR. We will come back to this point in section \ref{KS}.

If one is only interested in correlation functions for infinite system size at
finite temperatures or correlation functions for finite systems at zero
temperature, one can make use of the conformal invariance of the field theory.
It is then sufficient to calculate the correlation at zero temperature and
infinite system size and use a conformal mapping from the complex plane onto a
cylinder with the circumference corresponding either to inverse temperature or
system size. Here, we are, however, interested in the thermodynamics of finite
chains and we will therefore have to use an explicit mode expansion
 \begin{equation}
\label{FT.4}
\Phi(x=ja,t) = \sqrt{\frac{\pi}{8K}} +\sqrt{\frac{2\pi}{K}} S^z_{\rm tot} \frac{j}{N+1} 
+ \sum_{n=1}^\infty
\frac{\sin\l(\pi n j/(N+1)\r)}{\sqrt{\pi n}}\l(\e^{-i\frac{\pi n vt}{L+a}}b_n +
\e^{i\frac{\pi n vt}{L+a}}b_n^\dagger\r) 
\end{equation}
which incorporates the OBCs. Here $b_n$ is a bosonic annihilation operator.
Eq.~(\ref{FT.4}) is a discrete version of the mode expansions used in
\cite{EggertAffleck92,SirkerLaflorencie} with $x=ja$ becoming a continuous
coordinate for $a\to 0$, $N\to\infty$ with $L=Na$ fixed. Using this mode
expansion, the local observables respect the discrete lattice symmetry $j\to
N+1-j$ corresponding to a reflection at the central bond (site) for $N$ even
(odd), respectively. The sites $0$ and $N+1$ are added to model (\ref{eq1})
and we demand that the spin density vanishes at these sites. Therefore the
upper boundary for the integrals in (\ref{FT.1}) is $L+a$. The zero mode part
(first two terms of Eq.~(\ref{FT.4})) fulfills $\sum_j S^z_j \approx
\sqrt{\frac{K}{2\pi}}\int_0^{L+a} \partial_x\Phi \equiv S^z_{\rm tot}$ and the
oscillator part (last term of Eq.~(\ref{FT.4})) vanishes for $j=0,\, N+1$
as required.

For the free boson model the uniform part of the susceptibility can easily be
calculated and is given by \cite{EggertAffleck92}
\begin{equation} 
\label{scalechi}
\chi_0^{\rm uni} = -\frac{\partial^2 }{\partial h^2}\bigg|_{h=0}f_0 =
\frac{1}{NT} \frac{\sum_{S_z} S_z^2 
  e^{-\frac{\pi v}{K(L+a)T}S_z^2}}{\sum_{S_z} e^{-\frac{\pi v}{K(L+a)T}S_z^2}} 
= -\frac{1}{4NT}\frac{\partial^2}{\partial u^2}\bigg|_{u=0} \ln\theta\left(e^{-\frac{\pi v}{K(L+a)T}},u\right)
\end{equation}
Here $\theta(q,u)$ is the elliptic theta function of the third kind $\theta =
\theta_3(q,u) = \sum_{n=-\infty}^{\infty} q^{n^2} e^{i2 n u}$ for integer
$S_z$ (even $N$) and of the second kind $\theta = \theta_2(q,u) =
\sum_{n=-\infty}^{\infty} q^{(n+1/2)^2} e^{i(2 n+1) u}$ for half-integer $S_z$
(odd $N$). Note, that $\chi_0^{\rm uni}$ has a simple scaling form as a
function of $NT$. The lattice parameter $a$ appears here due to the OBCs. It
leads to a boundary correction which we will consider later. First, we set
$a\equiv 0$ in (\ref{scalechi}) and concentrate the following limiting cases
\begin{eqnarray}
\label{limits}
\chi_0^{\rm uni} &=& \l\{ \begin{array}{ll}
  \frac{2}{TN}\exp\l[-\frac{\pi v}{KLT}\r] & NT/v\to 0,\; \mbox{$N$ even} \\[0.2cm]
  \frac{1}{4TN} & NT/v\to 0,\; \mbox{$N$ odd} \\[0.2cm]
  \frac{K}{2\pi v} & NT/v\to\infty \; \; .
\end{array} \r.
\end{eqnarray}
The Curie-like divergence for $N$ odd is caused by the degeneracy of the
ground state $S^z_{\rm tot}=\pm 1/2$. At low temperatures the whole chain
therefore behaves like a single spin. For $N$ even, on the other hand, the
ground state is a singlet, $S^z_{\rm tot}=0$. At low temperatures the chain
becomes locked in this state leading to an exponentially small susceptibility.
In the thermodynamic limit, $ NT/v\to\infty$, the susceptibility within the
free boson approximation is just a constant.

The staggered part of the susceptibility, $\chi_j^{\rm st}$, for a finite
chain with OBCs has been calculated in \cite{SirkerLaflorencieNMR}. It is
given by
\begin{equation}
\label{StaggPart2}
\chi_j^{\rm st}
=-\frac{c}{T}\l(\frac{\pi}{N+1}\r)^{K/2}\!\!\!\!\!\!\frac{\eta^{3K/2}\l(\e^{-\frac{\pi
      v}{TL}}\r)}{\theta_1^{K/2}\l(\frac{\pi j}{N+1},\e^{-\frac{\pi
      v}{2TL}}\r)} \frac{\sum_{m} m\sin[2\pi m j/(N+1)]\e^{-\pi vm^2/(KLT)}}{\sum_{m}\e^{-\pi
    vm^2/(KLT)}} \; .
\end{equation}
Here $\eta(x)$ is the Dedekind eta-function and $\theta_1(u,q)$ the elliptic
theta-function of the first kind. The summation index $m$ runs over all
integers (half-integers) for $N$ even (odd), respectively. In the
thermodynamic limit, $N\to\infty$, we can simplify our result and obtain
\begin{equation}
  \label{StaggPart4}
\chi_j^{\rm st} = \frac{cK}{v}\frac{x}{\l[\frac{v}{\pi T}\sinh\l(\frac{2\pi T x}{v}\r)\r]^{K/2}}
\end{equation}
with $x=ja$. This agrees for the isotropic Heisenberg case, $K=1$, with the
result in \cite{EggertAffleck95}.  The amplitude $c$, first introduced in
Eq.~(\ref{FT.2}), can be determined with the help of the Bethe ansatz along the
lines of Ref.~\cite{LukyanovTerras}.  This leads to $c=\sqrt{A_z/2}$ with
$A_z$ as given in Eq.~(4.3) of \cite{LukyanovTerras}.  The formulas
(\ref{StaggPart2}) and (\ref{StaggPart4}) are therefore {\it parameter free}.

To find the boundary contributions, as for example the boundary free energy
defined in Eq.~(\ref{eq3}), one has to go beyond the free boson model
(\ref{FT.1}). When deriving the low-energy effective theory for the XXZ model,
one finds in addition to the free boson model (\ref{FT.1}) infinitely many
irrelevant terms. These terms either stem from band curvature (corrections to
the linear dispersions around the Fermi points) or from the interaction term.
For $\Delta$ close to $1$ the leading irrelevant term is given by
 \begin{equation}
   \delta H=\lambda\int_0^{L+a} dx \,\cos(\sqrt{8\pi K}\phi).
\label{int1}
\end{equation} 
It is the bosonized version of Umklapp scattering where two left moving
electrons get scattered to right movers or vice versa interchanging a
reciprocal lattice vector. This term becomes relevant for $\Delta >1$ and is
responsible for the opening of an excitation gap in this regime. For the
isotropic case, $\Delta =1$, Umklapp scattering is marginally irrelevant and
leads to important corrections to the results obtained for the free boson
model. Due to the integrability of the XXZ model the amplitude $\lambda$ of
the Umklapp term can be obtained exactly as well \cite{Lukyanov}. In
Ref.~\cite{SirkerLaflorencie,SirkerLaflorencie2} the free energy and the
susceptibility corrections to the free boson result to first order in the
Umklapp scattering have been calculated. For the susceptibility the following
correction was obtained
\begin{eqnarray}
\label{suscicorr}
\delta\chi_1^{\rm uni} &=&
\frac{2\lambda}{T^2}\l(\frac{\pi}{N}\r)^{2K}\!\!\eta^{6K}\l(\e^{-\frac{\pi
    v}{TL}}\r)  \int_0^{1/2} \!\!\!\!\!\! dy \,\frac{g_0\l(y,\e^{-\frac{\pi
      v}{KLT}}\r)}{\theta_1^{2K}\l(\pi y,\e^{-\frac{\pi v}{2TL}}\r)} 
\end{eqnarray}
with
\begin{eqnarray}
\label{suscicorr2}
g_0\l(y,q\r) &=& -\frac{\sum_{S_z} S_z^2\cos(4\pi S_z y) q^{S_z^2}}{\sum_{S_z}
  q^{S_z^2}} 
+\frac{\l(\sum_{S_z} \cos(4\pi S_z y) q^{S_z^2}\r)\l(\sum_{S_z}
  S_z^2  q^{S_z^2}\r)}{\l(\sum_{S_z}
  q^{S_z^2}\r)^2} \, .
\end{eqnarray}
In addition, there is also the boundary correction related to the parameter
$a$ in (\ref{scalechi}). Expanding in this parameter to lowest order we find
\begin{equation}
\label{NonU}
\delta \chi_2^{\rm uni} = \frac{\pi va}{KT^2L^3} g_2\l(\e^{-\frac{\pi
    v}{KLT}}\r) 
\end{equation}
where
\begin{equation}
\label{g2}
g_2\l(q\r) = \frac{\sum_{S_z} S_z^4  q^{S_z^2}}{\sum_{S_z}
  q^{S_z^2}} - \frac{\l(\sum_{S_z} S_z^2  q^{S_z^2}\r)^2}{\l(\sum_{S_z}
  q^{S_z^2}\r)^2} \, . 
\end{equation} 
$a$ plays the role of a lattice constant and its value can be determined for
$\Delta<1$ ($K>1$) by the Bethe ansatz and is given by
\cite{BortzSirker,SirkerBortzJSTAT}
\begin{equation}
\label{NonU3}
a=2^{-1/2}\sin\l[\pi K/(4K-4)\r]/\cos\l[\pi/(4K-4)\r] \; .
\end{equation}
The uniform part of the susceptibility of a finite chain is therefore given by
$\chi^{\rm uni}(L,T) =\chi_0^{\rm uni}(a\equiv 0) + \delta\chi_1^{\rm uni} +
\delta\chi_2^{\rm uni}$. The corrections to the uniform zeroth order
susceptibility $\chi_0^{\rm uni}(a\equiv 0)$ are important here because in the
thermodynamic limit they give the boundary susceptibility $\chi_B$. In this
limit $g_2\l(\e^{-\pi v/(KLT)}\r) \to K^2T^2L^2/(2\pi^2v^2)$ and therefore
\begin{equation}
\label{chi2}
\chi_{B,1}=\lim_{L\to\infty} L\chi_2^{\rm uni} =\frac{Ka}{2\pi v} \; .
\end{equation}  
This is just a constant contribution to the boundary susceptibility. Much more
important is the boundary contribution stemming from $\delta\chi_1^{\rm uni}$.
Here we find
\begin{equation}
  \chi_{B,2}=\lim_{L\to\infty} L\chi_1^{\rm uni} = - \lambda \l(\frac{K}{v}\r)^2
  B(K,1-2K)[\pi^2-2\psi'(K)]\l(\frac{2\pi T}{v}\r)^{2K-3}, 
\label{chi1}
\end{equation}
with $B(x,y)=\Gamma(x)\Gamma(y)/\Gamma(x+y)$, $\psi'(x)=d\psi(x)/dx$, and
$\psi(x)$ being the digamma function.  Note that for $1<K<3/2$
($1/2<\Delta<1$) the boundary spin susceptibility $\chi_{\rm B}$ shows a
divergent behavior $\sim 1/T^{3-2K}$ as temperature decreases.
\subsection{The isotropic point}
\label{iso}
At the isotropic point Umklapp scattering becomes marginal and simple
perturbation theory is no longer sufficient. In this case we have to replace
the Umklapp scattering amplitude by a running coupling constant $g(L,T)$ which
obeys a known set of renormalization group equations \cite{Lukyanov}
\begin{equation}
\label{coup_iso2}
1/g + \ln(g)/2 = \ln\l(\sqrt{2/\pi} e^{1/4+\gamma}\mbox{min}[L,v/T]\r) \; .
\end{equation} 
Here $\gamma\approx 0.577$ is Euler's constant and for the isotropic case
considered here, the spin velocity is $v=J\pi/2$. The uniform susceptibility
is then given by
\begin{equation}
\label{susci_iso}
\chi^{\rm uni}(N,T)=\chi_0^{\rm uni} + \delta\chi_1^{\rm uni} 
\end{equation}
where $K\to 1+g(L,T)/2$ in the exponentials of (\ref{scalechi}) and
$\lambda\to g(L,T)/4$ in (\ref{suscicorr}). In this case the parameter $a$ in
(\ref{scalechi}) is not determined by (\ref{NonU3}) and has to be used as a
fitting parameter.

In the thermodynamic limit we can again split the susceptibility
(\ref{susci_iso}) into a bulk and a boundary part. For the bulk susceptibility
this yields the result first derived by Lukyanov \cite{Lukyanov}
\begin{equation}
\label{bulk_thermo}
\chi_{\rm bulk} =
\frac{1}{\pi^2}\l(1+\frac{g(T)}{2}+\frac{3g^3(T)}{32}+\frac{\sqrt{3}}{\pi}T^2\r) \; .
\end{equation}
Here we have also added the $g^3$ correction from Umklapp scattering as well
as a $T^2$-term which stems from irrelevant operators with scaling dimension
$4$ describing band curvature. The running coupling constant $g(T)$ is given
by (\ref{coup_iso2}) with $L=\infty$. In the inset of Fig.~\ref{fig_ChiB}
this formula is displayed in comparison to the numerical results. For the
boundary susceptibility, on the other hand, we find
\begin{equation}
\label{chiB_iso3}
\chi_B = \frac{a}{\pi^2}+\frac{g}{12T}+\frac{g^2}{8T} \l(0.66 -
\frac{\Psi''(1)}{\pi^2}\r) + \mathcal{O}(g^3) \, .
\end{equation}
where we included second order corrections in $g$ as derived in
\cite{SirkerLaflorencie2}. The comparison of this formula with $a=1.5$ and
numerical results is shown in the main panel of Fig.~\ref{fig_ChiB}.

Finally, also the result for the staggered part of the susceptibility given in
Eq.~(\ref{StaggPart2}) has to be modified at the isotropic point. Here we find
\cite{SirkerLaflorencieNMR}
\begin{equation}
\label{StaggPart2_iso}
\chi_j^{\rm st}
=-\frac{1}{(2\pi^3 \tilde{g})^{1/4}T}\l(\frac{\pi}{N+1}\r)^{1/2}\!\!\!\!\!\!\frac{\eta^{3/2}\l(\e^{-\frac{\pi
      v}{TL}}\r)}{\theta_1^{1/2}\l(\frac{\pi j}{N+1},\e^{-\frac{\pi
      v}{2TL}}\r)} \frac{\sum_{m} m\sin[2\pi m j/(N+1)]\e^{-\pi vm^2(1-\tilde{g})/(LT)}}{\sum_{m}\e^{-\pi
    vm^2(1-\tilde{g})/(LT)}} \; .
\end{equation}
Now the running coupling constant $\tilde{g}$ also depends on the distance of
site $j$ from the boundary and is given by
\begin{equation}
\label{coup_iso21}
1/\tilde{g} + \ln(\tilde{g})/2 = \ln\l(\mbox{min}[C_0 x,C_0(L-x),\sqrt{\pi/2}\e^{1/4+\gamma}/T]\r) 
\end{equation} 
where the constant $C_0$ is not known and has to be used as a fitting
parameter. Note, however, that for low temperatures and $x,L-x\gg 1$ the value
of $C_0$ becomes irrelevant and our result for $\chi_j$ therefore again
parameter-free. If we consider the thermodynamic limit of
Eq.~(\ref{StaggPart2_iso}) we find
\begin{equation}
\label{StaggPart4_iso}
\chi_j^{\rm st} =
\l(\frac{2^3}{\pi^7\tilde{g}}\r)^{1/4}\frac{1}{1-\tilde{g}}\frac{x}{\sqrt{\frac{1}{2
      T}\sinh\l(4Tx\r)}} \; .
\end{equation}
A comparison of this formula with numerical data is shown in
Fig.~\ref{fig_chain}. The agreement is good but not perfect. The main problem
here is that the temperature and the length scale set by the distance from the
boundary are competing. The renormalization group equations, however, cannot
be solved with both scales present. The formula (\ref{coup_iso21}) is derived
in the limit when only one of those scales matter and the corrections can be
significant if this is not the case.

\section{The averaged susceptibility and a comparison with experimental data}
\label{Chi}
In an actual crystal we have imperfections and impurities which limit the
length of a spin chain segment. It is important to emphasize again that any
weakening of a link is a relevant perturbation in the renormalization group
sense. It is therefore expected to be a good approximation to assume that at
low temperatures we have spin chain segments with open boundary conditions. As
long as we do not know in detail how these defects are distributed it seems to
be reasonable to assume a Poisson distribution, i.e., the probability of
having an impurity at site $j$ is not influenced by the other impurities. We
might, however, expect this assumption to break down for large impurity
concentrations where some sort of impurity order might set in. Using a Poisson
distribution we have a normalized probability $P(N)=p(1-p)^N$ to find a chain
segment with $N$ sites if the impurity concentration is $p$. Any measurement
then corresponds to taking an average over this ensemble of chain segments.
For the susceptibility, for example, we find
\begin{equation}
\label{Susci}
\chi_p =p^2\sum_N N(1-p)^N \chi(N) \; .
\end{equation}  
Note that in a bulk measurement only the uniform part of the susceptibility
contributes. The staggered part cancels out. Using the formula
(\ref{susci_iso}) we can immediately calculate the average susceptibility
$\chi_p$. It is, however, very useful and instructive to derive a much simpler
approximate formula. To this end, we notice that we have two different regimes
for a finite spin chain. If the temperature is larger than the finite size
gap, $T/J> 1/N$, we can approximate $\chi(N)\approx \chi_\bulk +\chi_B/N$. If,
one the other hand, $T/J< 1/N$ we expect to see more or less the ground state
properties of the finite chain. This means, according to Eq.~(\ref{limits}),
that there will be no contribution when the chain length is even while
$\chi(N)\approx 1/(4TN)$ if the chain length is odd. We can therefore
introduce a length $N_c=\gamma J/T$ where this crossover occurs. Here $\gamma$
is a parameter which we expect to be of order $1$. For the average
susceptibility we can therefore write
\begin{eqnarray}
\label{fit_formula}
 \chi_p &\approx& \frac{p^2}{4T}\sum_{N\; \mbox{\tiny odd}}^{N_c}(1-p)^N +
p^2\sum_{N=N_c}^\infty (N\chi_\bulk+\chi_B) (1-p)^N \\
 &=& \frac{p}{4T}\frac{1-p}{2-p}\l(1-(1-p)^{\gamma/T}\r)+(1-p)^{\gamma/T}\l[\l(1-p+\frac{p\gamma}{T}\r)\chi_\bulk+p\chi_B\r] \nn
\end{eqnarray}
where $\chi_{\rm bulk}$ is given by Eq.~(\ref{bulk_thermo}) and $\chi_B$ by
Eq.~(\ref{chiB_iso3}).

\begin{figure}[t!]
\begin{center}
\includegraphics*[width=0.7\columnwidth]{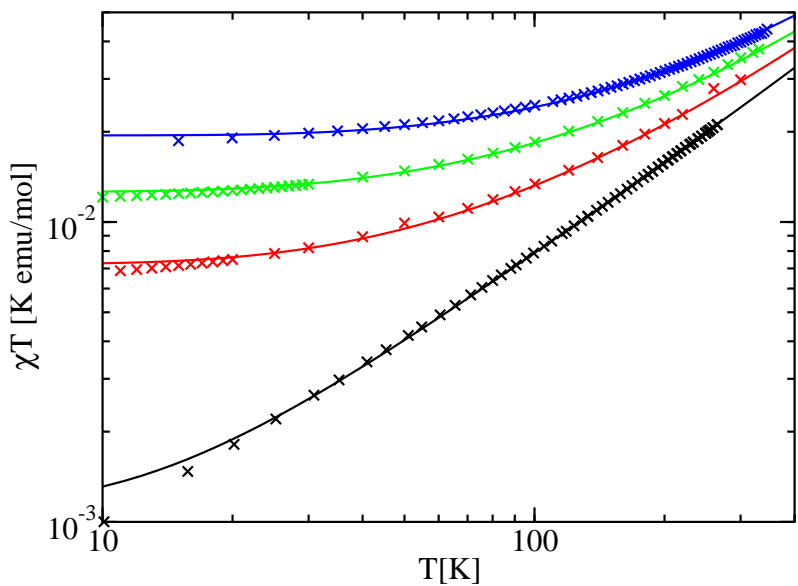}
\end{center}
\caption{Measured susceptibility $T(\chi-\chi_{\rm const})$ for
  Sr$_2$Cu$_{1-x}$Pd$_x$O$_{3+\delta}$ with impurity concentrations
  $x=0\%,0.5\%,1\%,3\%$ (crosses from bottom to top) from Ref.~\cite{Kojima}.
  Here a constant $\chi_{\rm const})$ as given in table \ref{tab1} has been
  subtracted from the experimental data. Subsequent curves are shifted by
  $5\times 10^{-3}$. For comparison theoretical results according to
  (\ref{fit_formula}) with $\gamma=1$ are shown with an effective impurity
  concentration $p$ as given in table \ref{tab1}.}
\label{Exp_fit}
\end{figure}

An interesting experiment has been performed by Kojima {\it et al.}
\cite{Kojima} where Palladium has been doped into the spin chain compound
Sr$_2$CuO$_3$. The Pd ions then replace the Cu ions, act as non-magnetic
impurities, and cut the chain into finite segments. A complication in the
analysis of this experiment occurs because even the undoped Sr$_2$CuO$_3$
samples already have a substantial amount of chain breaks. It is believed that
this is mainly a consequence of excess oxygen, i.e., we are in reality dealing
with Sr$_2$CuO$_{3+\delta}$. In a simple picture, an excess oxygen ion pulls
two electrons out of the copper chain. If these holes are relatively immobile
this also corresponds to a chain break. In the comparison of the experimental
data for Sr$_2$Cu$_{1-x}$Pd$_x$O$_{3+\delta}$ and formula (\ref{fit_formula})
shown in Fig.~\ref{Exp_fit} we therefore use $p$ as an effective impurity
concentration incorporating both the chain breaks due to excess oxygen and due
to the non-magnetic Pd ions.
Furthermore, we have subtracted a constant contribution $\chi_{\rm const}$
from the experimental data which is expected to be present due to core
diamagnetism and Van Vleck paramagnetism. The values for $p$ and $\chi_{\rm
  const}$ which yield the best fit of the experimental data are shown in table
\ref{tab1}.
\begin{table}[htbp]
  \caption{Concentration $x$ of Pd ions in experiment compared to chain break
    concentration $p$ and constant contribution $\chi_{\rm const}$ yielding the best
    theoretical fit. The first line corresponds to the ``as grown'' sample of Sr$_2$CuO$_{3+\delta}$ from
    Ref.~\cite{MotoyamaEisaki}.}
\label{tab1}
\begin{center}
\vspace*{0.4cm}
\begin{tabular}{c|c|c} 
$x$ (Exp.) & $p$ (Theory) & $\chi_{\rm const}$ [emu/mol] \\
\hline\hline\\[-0.3cm]
$0.0$ & $0.006$ & $-7.42\times 10^{-5}$ \\
$0.005$ & $0.012$ & $-7.7\times 10^{-5}$ \\
$0.01$ & $0.014$ & $-7.5\times 10^{-5}$ \\
$0.03$ & $0.024$ & $-7.5\times 10^{-5}$
\end{tabular}       
\end{center}                                                
\end{table}
For Pd concentrations of $x=0.5\%$ and $x=1\%$ the obtained values for $p$ are
consistent with the picture of having a certain amount of chain breaks due to
excess oxygen already in the undoped compound. For $x=3\%$, however, this
picture seems to fail. Reasons for this could be either on the experimental
side (perhaps not all Pd ions really go into the sample, or some go in
interstitially) or in the theoretical description. If, for example, the Pd
ions tend to cluster above a certain concentration, then our assumption of a
Poisson distribution becomes incorrect.

\section{The Knight shift and the role of interchain couplings}
\label{KS}
As already mentioned, the staggered part of the susceptibility cannot be
observed in a bulk measurement. This, however, is possible by NMR because here
the resonance frequency gets shifted by the {\it local} magnetic field. This
so-called Knight shift $K_j$ can therefore be directly related to the local
susceptibility. For a chain of length $N$ it is given by
\begin{equation}
\label{K1}
K^{(N)}_j=(\gamma_e/\gamma_n)\sum_{j'}A^{j-j'}\chi^{(N)}_{j'}
\end{equation} 
where $\gamma_e$ ($\gamma_n$) is the electron (nuclear) gyromagnetic ratio,
respectively and $A_{j-j'}$ the hyperfine coupling tensor. It is usually
sufficient to take only $A^0$ and $A^{\pm 1}$ into account because of the
short-range nature of the hyperfine interaction.

\begin{figure}[t!]
\begin{center}
\includegraphics*[width=0.7\columnwidth]{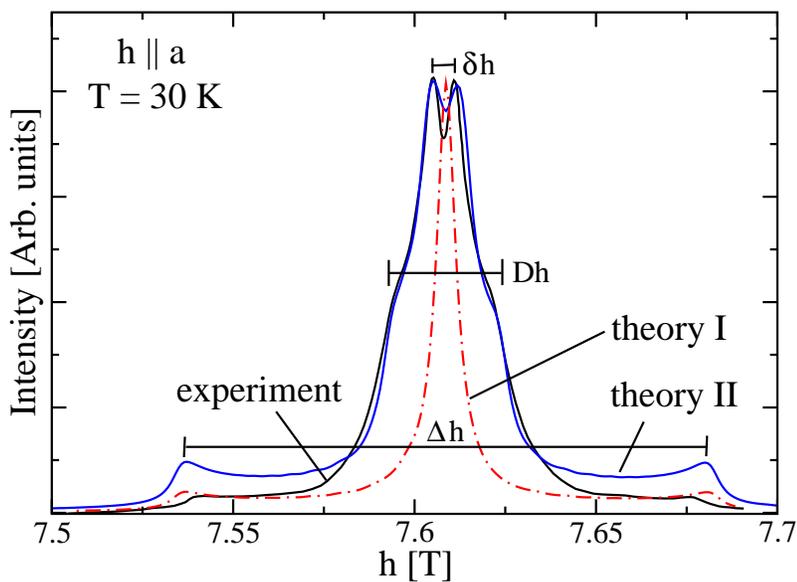}
\end{center}
\caption{NMR spectrum for Sr$_2$CuO$_3$ at $T=30$ K taken from
  \cite{TakigawaMotoyama}. In comparison the theoretical results for a single
  chain with a Poisson distribution of chain breaks (theory I) and for the
  case where also interchain couplings ($J_\perp =5$ K) are taken into account
  (theory II) are shown. In both cases $p=5\times 10^{-4}$, $\Gamma=4\times
  10^{-4}$, and $h^0_{\rm res} =7.598$ T. The material-dependent parameters are
  given in the text.}
\label{Fig_NMR}
\end{figure}

To compare with experiment, we assume again a Poisson distribution of chain
breaks so that the measured Knight shift is given by an average over all
possible chain lengths. Furthermore, each site in a chain of length $N$ gives
a different Knight shift according to (\ref{K1}). If we assume that each of
these Knight shifts has a Lorentzian lineshape with width $\Gamma$ we find for
the NMR spectrum
\begin{equation}
\label{Ks2}
 P(K) = \frac{\Gamma}{\pi}\!\!\sum_{N=1}^\infty
\frac{p(1-p)^{N-1}}{N}\!\sum_{j=1}^N\frac{1}{(K-K^{(N)}_j)^2+\Gamma^2} \; .
\end{equation}
Using the results for the uniform (\ref{susci_iso}) and the staggered part
(\ref{StaggPart2_iso}) we can now immediately calculate the NMR spectrum for
an ensemble of isotropic Heisenberg chain segments and compare to experimental
data for Sr$_2$CuO$_3$ obtained by Takigawa {\it et al.}
\cite{TakigawaMotoyama,BoucherTakigawa}. We use $J=2200$ K as exchange
constant in (\ref{eq1}) and hyperfine coupling constants
$A_c^0/(2\hbar\gamma_n) \approx -13$ T, $A_{ab}^0/(2\hbar\gamma_n) \approx 2$
T, and $A^1/(2\hbar\gamma_n) \approx 4$ T \cite{MonienPines}. Here the index
denotes the magnetic field direction. We calculate the spectrum as a function
of $h=(1+K)h_{\rm res}^0$ and use $h^0_{\rm res}=7.598$ T. For the resonance
field $\nu=86$ MHz used in the experiment \cite{TakigawaMotoyama} this is
consistent with $h_{\rm res}^0=\nu/\gamma_n$ where $\gamma_n\approx 11.3$
MHz/T \cite{AbragamBleaney}. A comparison of formula (\ref{Ks2}) for this set
of material-dependent parameters with experiment is shown in
Fig.~\ref{Fig_NMR}.
Here the impurity concentration $p$ and the Lorentzian linewidth $\Gamma$ have
been used as fitting parameters. Note that the sample used for the NMR
experiment has been annealed thus dramatically reducing the amount of excess
oxygen and the associated chain breaks compared to the sample used for the
susceptibility measurement shown in Fig.~\ref{Exp_fit}. The theory predicts a
central peak corresponding to the bulk susceptibility value and two shoulders
with separation $\Delta h$ (see curve 'theory I' in Fig.~\ref{Fig_NMR}) which
are caused by the maxima of the local susceptibility (see, for example,
Fig.~\ref{fig_chain}).  Theoretically, we find that $\Delta h\sim h^0_{\rm res}
\sqrt{v/T}\ln^{1/4}(v/T)$, i.e., the separation of the shoulders increases
$\sim 1/\sqrt{T}$ with decreasing temperature. This is in agreement with
experimental findings \cite{TakigawaMotoyama}. The central peak, however, has
a much more complex structure than predicted. The peak is split (feature
$\delta h$ in Fig.~\ref{Fig_NMR}) and has prominent shoulders (feature $D h$
in Fig.~\ref{Fig_NMR}). These features are only observed in experiment at
temperatures $T\lesssim 60$ K, whereas at higher temperatures the theoretical
prediction agrees well with experimental data (not shown).

To explain the additional features in the NMR spectra at $T\lesssim 60$ K we
have to take the interchain coupling into account. Sr$_2$CuO$_3$ shows
three-dimensional N\'eel ordering at $T_N\sim 5$ K. According to the usual
chain mean field argument the interchain coupling should therefore also be of
order $J_\perp\sim5$ K. This estimate is consistent with band structure
calculations \cite{RosnerEschrig}. As we have already demonstrated numerically
in Fig.~\ref{fig_ladder}, a ladder-like interchain coupling as present in
Sr$_2$CuO$_3$ leads to a reflection of susceptibility oscillations. Therefore
another typical Knight shift is expected to be present in the NMR spectra,
related to the maxima of the reflected oscillations. As long as $J_\perp/T \ll
1$ we can calculate this effect perturbatively. Starting from the definition
of the local susceptibility given in Eq.~(\ref{eq2}) we obtain to first order
\begin{equation}
\label{reflection0}
\chi_{j,1}^{(1)} =-\frac{J_\perp}{T^2}\sum_k\underbrace{\langle S^z_{j,1}S^z_{k,1}\rangle}_{G^{zz}_1(j-k)}
\underbrace{\langle S^z_{tot,2}S^z_{k,2}\rangle}_{\chi_{k,2}} \; .
\end{equation}
Here, the lower index $1$ stands for the infinitely long chain without chain
breaks whereas the lower index $2$ denotes the chain with a chain break at
site $k=0$. Both, the two-point correlation $G^{zz}_1(j-k)$ and the local
susceptibility $\chi_{k,2}$ have a uniform and a staggered part in the low
temperature limit. We therefore find $\chi_{j,1}^{(1)} \approx \chi_{j,1}^{\rm
  uni (1)}+ (-1)^j\chi_{j,1}^{\rm st (1)}$ with $\chi_{j,1}^{\rm uni (1)} =
-J_\perp \chi_1^{\rm uni (0)}\chi_2^{\rm uni (0)}/T$ and
\begin{equation}
\label{reflection}
\chi_{j,1}^{\rm st (1)} = -\frac{J_\perp}{T}(-1)^j \sum_k \chi_{k,2}^{\rm st} G^{zz,
  st}_1(j-k) \, .
\end{equation} 
Here Eq.~(\ref{StaggPart4}) has to be used for $\chi_{k,2}^{\rm st}$ while $G^{zz,
  st}_1(j-k) =\langle S^z_jS^z_k\rangle^{\rm st} = c^2/\l[\frac{v}{\pi
  T}\sinh(\frac{\pi T}{v}|j-k|)\r]^{K}$ is the staggered part of the bulk
two-point correlation function. In Fig.~\ref{fig_ladder}, formula
(\ref{reflection}) is compared with the numerical result and good agreement is
found.  

If we take these reflections into account, then also the additional features
in the NMR spectra are explained as shown in Fig.~\ref{Fig_NMR} (curve 'theory
II'). In particular, the shoulders with separation $Dh$ directly correspond to
the maxima of the reflected susceptibility oscillations. The splitting of the
peak $\delta h$ is of different origin. It is in fact not a splitting but
rather a loss of intensity at the value which corresponds to the bulk
susceptibility. The oscillations and reflected oscillations spread over the
entire crystal at low temperatures so that there are simply no sites left
which show bulk behavior. Rather interestingly, the peak usually associated
with the bulk susceptibility value therefore turns into a dip at low
temperatures due to the presence of chain breaks and interchain couplings.

Finally, we also want to shed some light on the role played by the geometry of
the interchain couplings. To this end, we consider an interchain coupling as
shown in Fig.~\ref{fig_geom}(c). Such a coupling is realized in SrCuO$_2$ with
a ferromagnetic $J_\perp \sim [-0.1\, J,-0.3\, J]$
\cite{MotoyamaEisaki,RosnerEschrig}. Using again first order perturbation
theory we find in this case
\begin{equation}
\label{reflection0_SrCuO2}
\chi_{j,1}^{(1)} =-\frac{J_\perp}{T^2}\sum_k\underbrace{\langle S^z_{j,1}S^z_{k,1}\rangle}_{G^{zz}_1(j-k)}
\underbrace{\langle S^z_{tot,2}(S^z_{k-1,2}+S^z_{k,2})\rangle}_{\chi_{k-1,2}+\chi_{k,2}} \; .
\end{equation}
Separating this into a uniform and a staggered part we find $\chi_{j,1}^{\rm uni (1)} =
-2J_\perp \chi_1^{\rm uni (0)}\chi_2^{\rm uni (0)}/T$ and
\begin{equation}
\label{reflection_SrCuO2}
\chi_{j,1}^{\rm st (1)} = -\frac{J_\perp}{T}(-1)^j \sum_k
\underbrace{(\chi_{k,2}^{\rm st}-\chi_{k-1,2}^{\rm st})}_{\approx 0} G^{zz,
  st}_1(j-k) \, .
\end{equation} 
Therefore no reflections will occur in this case to first order in
perturbation theory consistent with the numerical results shown in
Fig.~\ref{fig_zigzag}. This also means that an NMR spectrum for SrCuO$_2$
would not show any shoulders associated with the coupling to the nearest
neighbor chain. Reflections in chains further away might, however, be still
possible which then would again lead to additional structures in the NMR
spectra at low temperatures. 

\section{Summary and Conclusions}
We have investigated here how chain breaks and interchain couplings affect the
physical properties of spin chain compounds. A weak coupling of two Heisenberg
chains is an irrelevant perturbation in the renormalization group sense
whereas the weakening of a bond in an otherwise homogenous chain is relevant.
Open boundary conditions are therefore the stable fix point. Due to this
reasoning it is expected that a wide class of perturbations like impurities or
dislocations present in any real compound can be effectively described at low
energies as a chain break. An experimental measurement then corresponds to
taking an average over an ensemble of finite chain segments with open
boundaries.

By combining a low-energy effective field theory with Bethe ansatz we have
derived parameter-free formulas for the thermodynamics of finite spin-$1/2$
Heisenberg chains with open boundary conditions. Particular emphasis was put
on a calculation of the susceptibility. Due to the broken translational
invariance there exists a site-dependent staggered susceptibility in addition
to the uniform site-independent part. However, even the uniform part is
affected by the open boundary conditions in the sense that a surface
contribution arises which is not present for periodic boundary conditions.

We have shown that susceptibility measurements on Sr$_2$CuO$_3$ doped with
non-magnetic Pd ions are well described by the theory presented here. One of
the complications arising for this compound is, however, that even the undoped
sample apparently already has a relatively large amount of chain breaks which
are believed to be caused by excess oxygen. The impurity concentration used in
the theory to fit the experimental data therefore differs significantly from
the nominal Pd concentration.  The emerging picture nevertheless seems to be
consistent - at least at low impurity concentrations - with a fixed
concentration of additional chain breaks already present in the undoped
sample. It would certainly be of some value to obtain experimental data for
this or some other spin-chain compound where the impurity concentration is
well controlled and a direct comparison with theory therefore possible.

Furthermore, we used the field theory to calculate NMR spectra.  Importantly,
the Knight shift is proportional to the {\it local} susceptibility so that the
staggered site-dependent part of the susceptibility, which cannot be observed
in a bulk susceptibility measurement, becomes observable. As has already been
shown previously \cite{TakigawaMotoyama}, the staggered susceptibility leads
to a broad background in the NMR spectra with edges caused by the maxima of
the staggered susceptibility. For Sr$_2$CuO$_3$ it has, however been found
that the NMR spectra at low temperatures show puzzling additional features
which have been ascribed in \cite{BoucherTakigawa} to a coupling to phonons.
Here we have shown that these features quite naturally arise in a spin-only
model if the known interchain couplings are taken into account. It is also
important to note that the geometry of the interchain couplings is crucial. As
a specific example we considered in addition to the ladder-like interchain
coupling relevant for Sr$_2$CuO$_3$ also the zigzag-like interchain coupling
realized in SrCuO$_2$. In the latter case no additional structures in the NMR
spectra related to this interchain coupling will occur. NMR experiments on
spin chain compounds are therefore not only helpful to study impurity effects,
they can also be used to investigate the geometry and strength of interchain
couplings.

\section*{Acknowledgments}
I want to thank all my collaborators on this and related topics, in
particular, Ian Affleck, Michael Bortz, Sebastian Eggert, Andreas Kl\"umper,
and Nicolas Laflorencie.



%
%


\end{document}